\documentclass[titlepage,11pt]{article}

\usepackage{graphicx}
\usepackage[myheadings]{fullpage}
\usepackage{pmetrika}
\usepackage{natbib}
\usepackage{submit}
\usepackage[T1]{fontenc}
\usepackage{parskip}
\usepackage{latexsym,amsmath,amssymb}
\usepackage{comment}
\usepackage{times}
\usepackage{zi4}
\usepackage{subfigure}
\usepackage{hyperref}
\usepackage{float}
\usepackage{enumerate}
\usepackage{bm,bbm}
\usepackage{listings}

\usepackage{xcolor}
\usepackage{multirow,booktabs,setspace,tabularx,ragged2e,threeparttable}
\newcolumntype{L}{>{\RaggedRight\hangafter=1\hangindent=0em}X}
\newcolumntype{C}{>{\centering\arraybackslash}X}
\usepackage{tikz}
\usepackage[ruled,vlined]{algorithm2e}
\usepackage{etoolbox}
\AtBeginEnvironment{lstlisting}{\singlespacing}

\newcommand{\E}{\mathbb{E}}
\newcommand{\Var}{\mathrm{Var}}

\begin{document}

\begin{titlepage}

\title{Robust Standard Errors for Bayesian Posterior Functionals via the Infinitesimal Jackknife}

\author{Nanyu Luo, Feng Ji}

\affil{University of Toronto}

\vspace{\fill}\centerline{\today}\vspace{\fill}

\linespacing{1}
\contact{
The authors have no relevant financial or non-financial interests to disclose.
The authors have no conflicts of interest to declare that are relevant to the content of this article.
All authors certify that they have no affiliations with or involvement in any organization or entity
with any financial interest or non-financial interest in the subject matter or materials discussed in
this manuscript.
The authors have no financial or proprietary interests in any material discussed in this article.\newline
\indent{\it Address for correspondence}: Feng Ji,
University of Toronto, 252 Bloor St W, Toronto, ON M5S 1V6, Canada.
{\sf E-mail: f.ji@utoronto.ca}}

\end{titlepage}

\setcounter{page}{2}
\vspace*{2\baselineskip}

\RepeatTitle{Robust Standard Errors for Bayesian Posterior Functionals via the Infinitesimal Jackknife}\vskip3pt

\linespacing{1.5}
\abstracthead
\begin{abstract}
Quantitative research in the social and behavioral sciences relies heavily on nonlinear posterior functionals such as indirect effects, standardized coefficients, effect sizes, intraclass correlations, and multilevel variance-explained measures. The posterior standard deviation (PostSD) is the default uncertainty summary for these quantities, yet it presupposes a correctly specified model. When the working model is wrong, as is common with behavioral data that exhibit heavy tails and heteroskedasticity, PostSD can severely underestimate the frequentist standard error. The nonparametric bootstrap offers robustness but requires repeated MCMC refits, while the delta method demands a separate analytic gradient derivation for every new functional. The infinitesimal jackknife standard error \citep[IJSE,][]{giordano2019swiss,GiordanoBroderick2023} sidesteps both limitations: it approximates the bootstrap variance through influence functions computed from a single MCMC run, applies to any posterior functional without modification, and requires no analytic derivatives. We discuss the use the IJSE methodology at both the observation level and the cluster level and evaluate it through four simulation studies covering six functionals from mediation analysis, ANOVA, and multilevel modeling, which are commonly used in the social and behavioral sciences. Under misspecification, PostSD substantially underestimated the true standard error across all settings, whereas IJSE closely tracked the bootstrap at a fraction of the computational cost. Under correct specification all three methods agreed, confirming that IJSE introduces no distortion when the model is right. These results show IJSE as a practical, general-purpose tool for robust uncertainty quantification in Bayesian workflows throughout the social and behavioral sciences.

\end{abstract}\vspace{\fill}\pagebreak

\setlength{\abovedisplayskip}{6pt}
\setlength{\belowdisplayskip}{6pt}
\setlength{\abovedisplayshortskip}{4pt}
\setlength{\belowdisplayshortskip}{4pt}
\setlength\jot{4pt}%

\section{Introduction}

Bayesian methods are now widely adopted in the social and behavioral sciences for their capacity to incorporate prior knowledge, handle complex hierarchical structures, and produce coherent probabilistic inference \citep{GelmanBDA3}. In many applied settings the scientific quantity of interest is not a raw model parameter but a functional of the parameter vector $\bm{\theta}$. For example, mediation analysis relies on products of regression coefficients \citep{BaronKenny1986,MacKinnon2008}; standardized indirect effects further divide by a model-implied standard deviation \citep{PreacherKelley2011,LachowiczPreacherKelley2018}; ANOVA effect sizes such as $\eta^2$ involve ratios of sums of squares \citep{Lakens2013}; and multilevel models give rise to the intraclass correlation coefficient and the marginal and conditional $R^2$, both defined through ratios of variance components \citep{ShroutFleiss1979,RaudenbushBryk2002,NakagawaSchielzeth2013}. These functionals are among the most frequently reported quantities in psychology, education, and public health, yet obtaining reliable standard errors for them remains a nontrivial problem. Given posterior draws $\bm{\theta}^{(1)},\ldots,\bm{\theta}^{(T)}$ from Markov chain Monte Carlo (MCMC), the conventional approach estimates $g(\bm{\theta})$ by the posterior mean $\bar{g} = T^{-1}\sum_t g(\bm{\theta}^{(t)})$ and summarizes uncertainty by the posterior standard deviation (PostSD).

PostSD is a valid standard error only when the parametric working model is correctly specified. Under correct specification the Bernstein--von Mises theorem guarantees that the posterior is asymptotically equivalent to the sampling distribution of the maximum likelihood estimator, so that PostSD coincides with the frequentist standard error and credible intervals double as confidence intervals \citep{vandervaart1998asymptotic,Rubin1984}. In practice, however, working models are almost always approximations. Behavioral data routinely exhibit heavier tails, heteroskedasticity, and distributional asymmetries that violate Gaussian assumptions \citep{Micceri1989,Wilcox2012}. Under such misspecification the posterior still concentrates around a pseudo-true parameter that minimizes the Kullback--Leibler divergence to the true distribution \citep{White1982,KleijnVanDerVaart2012}, but its spread reflects the model-based Fisher information $H$ rather than the true sampling variability captured by the sandwich form $H^{-1}JH^{-1}$ \citep{huber1967behavior,White1982,Muller2013}. The gap between $H^{-1}$ and $H^{-1}JH^{-1}$ does not vanish with increasing sample size, so PostSD can persistently underestimate the frequentist standard error, yielding credible intervals that are too narrow and coverage probabilities that fall well below nominal levels.

This concern is especially acute for functionals that depend on variance components. Standardized indirect effects, $\eta^2$, intraclass correlations, and conditional $R^2$ all place residual, between-group, or between-cluster variances in their numerators or denominators. A misspecified Gaussian likelihood produces an overly concentrated posterior for these variance parameters, because the Gaussian assumption treats every observation as equally informative regardless of actual tail behavior \citep{GiordanoBroderick2023}. The resulting underestimation of uncertainty propagates nonlinearly through the functional, making PostSD unreliable for precisely the quantities that applied researchers report most frequently.

Two established remedies exist in the frequentist toolkit, but each has significant limitations. The nonparametric bootstrap \citep{Efron1979,DavisonHinkley1997} yields a model-free standard error by resampling the data and refitting the model on each resample \citep{BollenStine1990,MacKinnon2004}. For Bayesian estimators this requires a full MCMC run on each of $B$ bootstrap samples, multiplying the computational cost by a factor of $B$. The delta method \citep{Oehlert1992} avoids resampling but requires the analyst to derive the analytic gradient $\nabla g(\bm{\theta})$ for each functional of interest. For simple functionals such as the product $ab$ in mediation \citep{Sobel1982} the derivatives are straightforward, but for standardized indirect effects, multilevel $R^2$, and other compound ratios the derivations become increasingly tedious and error-prone \citep{PreacherKelley2011,NakagawaSchielzeth2013,LachowiczPreacherKelley2018}. Each new functional demands its own algebraic treatment, and combining the delta method with a clustered sandwich estimator for multilevel data adds further complexity. In practice this per-functional derivation burden means that applied researchers often lack access to robust standard errors for precisely the quantities they wish to report.

The infinitesimal jackknife (IJ) offers an alternative that combines the robustness of the bootstrap with the single-fit efficiency of a closed-form estimator, while requiring no analytic derivatives. Originally introduced by \citet{Jaeckel1972}, the IJ approximates the jackknife variance through influence functions. \citet{giordano2019swiss} developed a general IJ framework for optimization-based estimators, and \citet{GiordanoBroderick2023} extended it to the Bayesian setting, showing that the influence of each observation on a posterior mean can be expressed as a posterior covariance between the observation's log-likelihood contribution and the functional of interest. The resulting IJ standard error (IJSE) can be computed entirely from a single set of MCMC draws at an additional cost of $O(NT)$, which is typically negligible relative to the MCMC computation itself. Because the influence proxies are empirical covariances between log-likelihood contributions and draws of $g(\bm{\theta}^{(t)})$, the same algorithm applies to any posterior functional without modification: once the MCMC output is in hand, switching from one functional to another amounts to changing a single line of code.

Despite these attractive properties and great potentials for many models and procedures in social and bahavioral sciences, IJSE has not been systematically discussed and evaluated for the types of posterior functionals most commonly encountered in social and behavioral research. The existing literature has focused on point-estimation sensitivity and model-checking applications \citep{giordano2019swiss} rather than on standard error estimation across a range of applied functionals. This paper fills that gap through four simulation studies of increasing complexity. We evaluate IJSE for the unstandardized and standardized indirect effects in linear mediation, the ANOVA effect size $\eta^2$ in a one-way between-subjects design, the intraclass correlation in a random-intercept model with a fixed-effect covariate, and the marginal and conditional $R^2$ from the same multilevel model. In each study the data-generating process features heavy-tailed errors and predictor-dependent heteroskedasticity that violate the Gaussian working model, reflecting the types of distributional departures documented as empirically common in behavioral data \citep{Micceri1989,Wilcox2012}. Across all settings we find that PostSD substantially underestimates the frequentist standard error, while IJSE closely tracks the nonparametric bootstrap at a fraction of the computational cost.

The remainder of the paper is organized as follows. Section~\ref{sec:background} reviews the relevant background on Bayesian posterior functionals, frequentist variance under misspecification, and the sandwich covariance matrix. Section~\ref{sec:ijse} introduces the IJSE methodology at both the observation and cluster levels, presents the computation algorithms, and compares computational costs with the bootstrap. Section~\ref{sec:sim1} examines linear mediation under both correct specification and misspecification, evaluating IJSE for the unstandardized and standardized indirect effects. Section~\ref{sec:sim3} evaluates IJSE for ANOVA effect sizes, Section~\ref{sec:sim4} considers the intraclass correlation in a multilevel model with a fixed-effect covariate, and Section~\ref{sec:sim_r2} extends the same multilevel framework to the marginal and conditional $R^2$.

\section{Background and Notation}
\label{sec:background}

\subsection{Bayesian Inference and Posterior Functionals}

Consider a parametric model $p(y_i \mid \bm{\theta})$ for independent observations $y_1, \ldots, y_N$, with prior $p(\bm{\theta})$. The posterior distribution is
\begin{equation}
p(\bm{\theta} \mid y_{1:N}) \propto p(\bm{\theta}) \prod_{i=1}^{N} p(y_i \mid \bm{\theta}).
\label{eq:posterior}
\end{equation}
In many applications, the quantity of interest is not $\bm{\theta}$ itself but a functional $g(\bm{\theta})$, such as a predicted probability, a causal effect, or a transformed parameter. Given MCMC draws $\bm{\theta}^{(1)}, \ldots, \bm{\theta}^{(T)}$ from \eqref{eq:posterior}, the standard Bayesian point estimate is the posterior mean
\begin{equation}
\bar{g} = \frac{1}{T} \sum_{t=1}^{T} g(\bm{\theta}^{(t)}),
\label{eq:post_mean}
\end{equation}
and uncertainty is typically summarized by the posterior standard deviation (PostSD),
\begin{equation}
\widehat{\mathrm{SE}}_{\mathrm{PostSD}} = \mathrm{sd}\bigl(\{g(\bm{\theta}^{(t)})\}_{t=1}^{T}\bigr).
\label{eq:postsd_def}
\end{equation}

\subsection{Frequentist Variance of Bayesian Estimators}

From a frequentist perspective, $\bar{g}$ is a point estimator whose sampling distribution depends on the data-generating process (DGP). Let $\bar{g}(y_{1:N})$ denote the posterior mean computed from data $y_{1:N}$. Under repeated sampling, the frequentist variance is
\begin{equation}
\Var_{\mathrm{freq}}(\bar{g}) = \E\left[(\bar{g} - \E[\bar{g}])^2\right],
\label{eq:freq_var}
\end{equation}
where the expectation is over the true DGP. In correctly specified models with large $N$, the Bernstein--von Mises theorem implies that the posterior concentrates around the true parameter, and PostSD approximates the frequentist standard error \citep{vandervaart1998asymptotic}.

\subsection{The Sandwich Variance Under Misspecification}
\label{sec:sandwich}

When the working model is misspecified, the posterior may concentrate around a pseudo-true parameter $\bm{\theta}^\star$ that minimizes Kullback--Leibler divergence to the true distribution \citep{White1982,KleijnVanDerVaart2012}. In this setting, the model-based variance $H^{-1}$ (inverse Fisher information) no longer equals the frequentist variance. Instead, the asymptotic variance takes the sandwich form
\begin{equation}
\Var_{\mathrm{freq}}(\hat{\bm{\theta}}) \approx H^{-1} J H^{-1},
\label{eq:sandwich}
\end{equation}
where $H$ is the expected Hessian of the log-likelihood and $J$ is the variance of the score function \citep{White1982,huber1967behavior}. When the model is correct, $J = H$ and the sandwich reduces to $H^{-1}$. Under misspecification, $J \neq H$ in general, and the model-based variance can differ substantially from the true frequentist variance.

For Bayesian estimators under misspecification, analogous results hold: PostSD may not reflect the true sampling variability of $\bar{g}$, potentially leading to miscalibrated credible intervals \citep{KleijnVanDerVaart2012,Muller2013}. This motivates variance estimation approaches that remain valid regardless of model correctness.

\section{The Infinitesimal Jackknife for Bayesian Functionals}
\label{sec:ijse}

This section introduces the infinitesimal jackknife standard error \citep{GiordanoBroderick2023} for Bayesian posterior functionals. Section~\ref{subsec:ijse_obs} presents the observation-level formulation for independent data. Section~\ref{subsec:ijse_cluster} extends the framework to clustered data, where the independent sampling units are clusters rather than individual observations. Sections~\ref{subsec:ijse_bootstrap} and~\ref{subsec:ijse_cost} then compare IJSE with the nonparametric bootstrap and discuss computational costs.

\subsection{IJSE for Bayesian Posterior Means: Observation Level}
\label{subsec:ijse_obs}

The IJ provides a computationally efficient approximation to the frequentist variance of an estimator by exploiting influence functions \citep{Jaeckel1972,efron1982jackknife}. For a functional $\bar{g}$ computed from data $y_{1:N}$, the influence of observation $i$ measures how much $\bar{g}$ would change if observation $i$ were upweighted infinitesimally. Let $\bar{g}(\bm{w})$ denote the estimator computed with observation weights $\bm{w} = (w_1, \ldots, w_N)$, where $w_i = 1$ corresponds to the original data. The empirical influence function is
\begin{equation}
I_i = N \cdot \left.\frac{\partial \bar{g}(\bm{w})}{\partial w_i}\right|_{\bm{w}=\bm{1}}.
\label{eq:influence}
\end{equation}
The IJ variance estimator is
\begin{equation}
\widehat{\Var}_{\mathrm{IJ}}(\bar{g}) = \frac{1}{N(N-1)} \sum_{i=1}^{N} (I_i - \bar{I})^2,
\label{eq:ij_var}
\end{equation}
where $\bar{I} = N^{-1}\sum_{i=1}^{N} I_i$. This approximates the variance of the leave-one-out jackknife without requiring $N$ refits \citep{efron1982jackknife}.

For Bayesian posterior means, \citet{GiordanoBroderick2023} showed that the influence function can be computed as a posterior covariance. Let $L_i^{(t)} = \log p(y_i \mid \bm{\theta}^{(t)})$ denote the log-likelihood contribution of observation $i$ at MCMC draw $t$. The influence of observation $i$ on the posterior mean $\bar{g}$ is approximated by
\begin{equation}
I_i \approx N \cdot \widehat{\mathrm{Cov}}_t\left(L_i^{(t)}, g(\bm{\theta}^{(t)})\right),
\label{eq:ij_cov}
\end{equation}
where $\widehat{\mathrm{Cov}}_t$ denotes the sample covariance across MCMC draws. This result enables computation of the IJ variance from a \emph{single} MCMC run, without any refitting. Given MCMC draws $\{\bm{\theta}^{(t)}\}_{t=1}^{T}$ from the posterior, IJSE is computed via Algorithm~\ref{alg:ijse}. The key quantities are the influence proxies
\begin{equation}
I_i = N \cdot \frac{\sum_{t=1}^{T} \tilde{L}_i^{(t)} \tilde{g}^{(t)}}{T - 1},
\label{eq:Ii_formula}
\end{equation}
where $\tilde{g}^{(t)} = g^{(t)} - \bar{g}$ are centered functional draws and $\tilde{L}_i^{(t)} = L_i^{(t)} - \bar{L}^{(t)}$ are log-likelihood contributions centered across observations. IJSE is then
\begin{equation}
\widehat{\mathrm{SE}}_{\mathrm{IJ}} = \sqrt{\frac{1}{N(N-1)} \sum_{i=1}^{N} (I_i - \bar{I})^2}.
\label{eq:ijse_formula}
\end{equation}

\begin{algorithm}[!htb]
\DontPrintSemicolon
\caption{IJSE from a single MCMC run}
\label{alg:ijse}
\KwIn{MCMC draws $\{\bm{\theta}^{(t)}\}_{t=1}^{T}$; data $\{y_i\}_{i=1}^N$; functional $g(\cdot)$}
\KwOut{$\widehat{\mathrm{SE}}_{\mathrm{IJ}}$}
\For{$t=1$ \KwTo $T$}{
  Compute $g^{(t)} \leftarrow g(\bm{\theta}^{(t)})$\;
  \For{$i=1$ \KwTo $N$}{
    Compute $L_i^{(t)} \leftarrow \log p(y_i \mid \bm{\theta}^{(t)})$\;
  }
}
Compute $\bar{g} \leftarrow T^{-1}\sum_t g^{(t)}$ and $\tilde{g}^{(t)} \leftarrow g^{(t)} - \bar{g}$\;
\For{$t=1$ \KwTo $T$}{
  Compute $\bar{L}^{(t)} \leftarrow N^{-1}\sum_j L_j^{(t)}$ and $\tilde{L}_i^{(t)} \leftarrow L_i^{(t)} - \bar{L}^{(t)}$\;
}
\For{$i=1$ \KwTo $N$}{
  Compute $I_i \leftarrow N \cdot (T-1)^{-1} \sum_t \tilde{L}_i^{(t)} \tilde{g}^{(t)}$\;
}
Compute $\bar{I} \leftarrow N^{-1}\sum_i I_i$ and $\widehat{\mathrm{SE}}_{\mathrm{IJ}} \leftarrow \sqrt{[N(N-1)]^{-1}\sum_i (I_i - \bar{I})^2}$\;
\Return $\widehat{\mathrm{SE}}_{\mathrm{IJ}}$\;
\end{algorithm}

\subsection{IJSE for Bayesian Posterior Means: Cluster Level}
\label{subsec:ijse_cluster}

In multilevel models, observations are nested within clusters, and the natural unit of independent sampling is the cluster rather than the individual observation. When $K$ clusters of size $m$ are drawn independently from a population, the posterior depends on the data through $K$ cluster-level contributions, and the IJ variance must be computed over these $K$ units rather than over all $N = Km$ observations.

Let $\bm{Y}_k = (Y_{1k}, \ldots, Y_{mk})$ denote the observations in cluster $k$, let $\bm{x}_k = (x_{1k}, \ldots, x_{mk})$ denote any cluster-member covariates, and let $U_k$ denote the cluster-specific random effect sampled jointly with the model parameters. The cluster-level log-likelihood contribution at posterior draw $t$ aggregates the random-effect density and all within-cluster observation densities:
\begin{equation}
L_k^{(t)} = \log p\!\left(U_k^{(t)} \mid \bm{\theta}^{(t)}\right) + \sum_{i=1}^{m} \log p\!\left(Y_{ik} \mid U_k^{(t)}, \bm{x}_k, \bm{\theta}^{(t)}\right),
\qquad k = 1, \ldots, K.
\label{eq:cluster_Lk}
\end{equation}
The influence proxies and variance estimator then follow \eqref{eq:Ii_formula}--\eqref{eq:ijse_formula} with the substitutions $N \to K$ and $L_i^{(t)} \to L_k^{(t)}$:
\begin{equation}
I_k = K \cdot \frac{\sum_{t=1}^{T} \tilde{L}_k^{(t)}\, \tilde{g}^{(t)}}{T-1},
\qquad
\widehat{\mathrm{SE}}_{\mathrm{IJ}} = \sqrt{\frac{1}{K(K-1)} \sum_{k=1}^{K} (I_k - \bar{I})^2},
\label{eq:cluster_ijse}
\end{equation}
where $\tilde{L}_k^{(t)} = L_k^{(t)} - K^{-1}\sum_{j=1}^{K} L_j^{(t)}$ and $\tilde{g}^{(t)} = g^{(t)} - \bar{g}$. The key distinction from the observation-level case is that the number of independent units entering the variance estimator is $K$, not the total number of observations $N = Km$. Consequently, the IJ approximation requires $K$ to be sufficiently large for the influence-function variance to stabilize.

Algorithm~\ref{alg:ijse_cluster} summarizes the procedure. Because the log-likelihood matrix $\{L_k^{(t)}\}$ is computed once and reused for every target functional, multiple variance-component quantities such as the ICC, marginal $R^2$, and conditional $R^2$ can be evaluated from the same MCMC output at negligible incremental cost.

\begin{algorithm}[!htb]
\DontPrintSemicolon
\caption{Cluster-level IJSE from a single MCMC run}
\label{alg:ijse_cluster}
\KwIn{MCMC draws $\{\bm{\theta}^{(t)}, U_1^{(t)}, \ldots, U_K^{(t)}\}_{t=1}^{T}$; clustered data $\{(\bm{Y}_k, \bm{x}_k)\}_{k=1}^K$; functional $g(\cdot)$}
\KwOut{$\widehat{\mathrm{SE}}_{\mathrm{IJ}}$}
\For{$t=1$ \KwTo $T$}{
  Compute $g^{(t)} \leftarrow g(\bm{\theta}^{(t)})$\;
  \For{$k=1$ \KwTo $K$}{
    Compute $L_k^{(t)} \leftarrow \log p(U_k^{(t)} \mid \bm{\theta}^{(t)}) + \sum_{i=1}^{m} \log p(Y_{ik} \mid U_k^{(t)}, \bm{x}_k, \bm{\theta}^{(t)})$\;
  }
}
Compute $\bar{g} \leftarrow T^{-1}\sum_t g^{(t)}$ and $\tilde{g}^{(t)} \leftarrow g^{(t)} - \bar{g}$\;
\For{$t=1$ \KwTo $T$}{
  Compute $\bar{L}^{(t)} \leftarrow K^{-1}\sum_j L_j^{(t)}$ and $\tilde{L}_k^{(t)} \leftarrow L_k^{(t)} - \bar{L}^{(t)}$\;
}
\For{$k=1$ \KwTo $K$}{
  Compute $I_k \leftarrow K \cdot (T-1)^{-1} \sum_t \tilde{L}_k^{(t)} \tilde{g}^{(t)}$\;
}
Compute $\bar{I} \leftarrow K^{-1}\sum_k I_k$ and $\widehat{\mathrm{SE}}_{\mathrm{IJ}} \leftarrow \sqrt{[K(K-1)]^{-1}\sum_k (I_k - \bar{I})^2}$\;
\Return $\widehat{\mathrm{SE}}_{\mathrm{IJ}}$\;
\end{algorithm}

\subsection{Relationship to Nonparametric Bootstrap}
\label{subsec:ijse_bootstrap}

The nonparametric (pairs) bootstrap \citep[NP;][]{Efron1979,DavisonHinkley1997} provides another approach to estimating the frequentist variance of $\bar{g}$. For $b = 1, \ldots, B$, resample $N$ observations with replacement, refit the model via MCMC, compute $\bar{g}_b^*$, and take
\begin{equation}
\widehat{\mathrm{SE}}_{\mathrm{NP}} = \mathrm{sd}\left(\{\bar{g}_b^*\}_{b=1}^B\right).
\label{eq:np_se}
\end{equation}
Both IJSE and the nonparametric bootstrap target the same quantity, the frequentist SE of $\bar{g}$, and should agree asymptotically. The key difference is computational: the bootstrap requires $B$ complete MCMC refits, while IJSE requires only one baseline fit plus $O(NT)$ covariance computations. For clustered data, the bootstrap resamples entire clusters with replacement, with the number of refits still equal to $B$, and IJSE replaces the $O(NT)$ cost with $O(KT)$.

\subsection{Computational Complexity Comparison}
\label{subsec:ijse_cost}

Let $C_{\mathrm{MCMC}}$ denote the cost of one MCMC run with $T$ iterations. The computational costs are:
\begin{itemize}
\item \textbf{PostSD}: $C_{\mathrm{MCMC}}$ (one MCMC run)
\item \textbf{IJSE}: $C_{\mathrm{MCMC}} + O(NT)$ (one MCMC run + covariance computation)
\item \textbf{Nonparametric bootstrap}: $B \cdot C_{\mathrm{MCMC}}^{\mathrm{boot}}$ ($B$ MCMC runs, possibly with shorter chains)
\end{itemize}
For typical settings with $B = 200$ and $C_{\mathrm{MCMC}}^{\mathrm{boot}} \approx 0.3 \cdot C_{\mathrm{MCMC}}$, the bootstrap is roughly $60\times$ more expensive than IJSE. This speedup is particularly valuable when MCMC is computationally intensive, such as in hierarchical models or when using gradient-based samplers.

\begin{figure}[!htb]
  \centering
  \includegraphics[width=\textwidth]{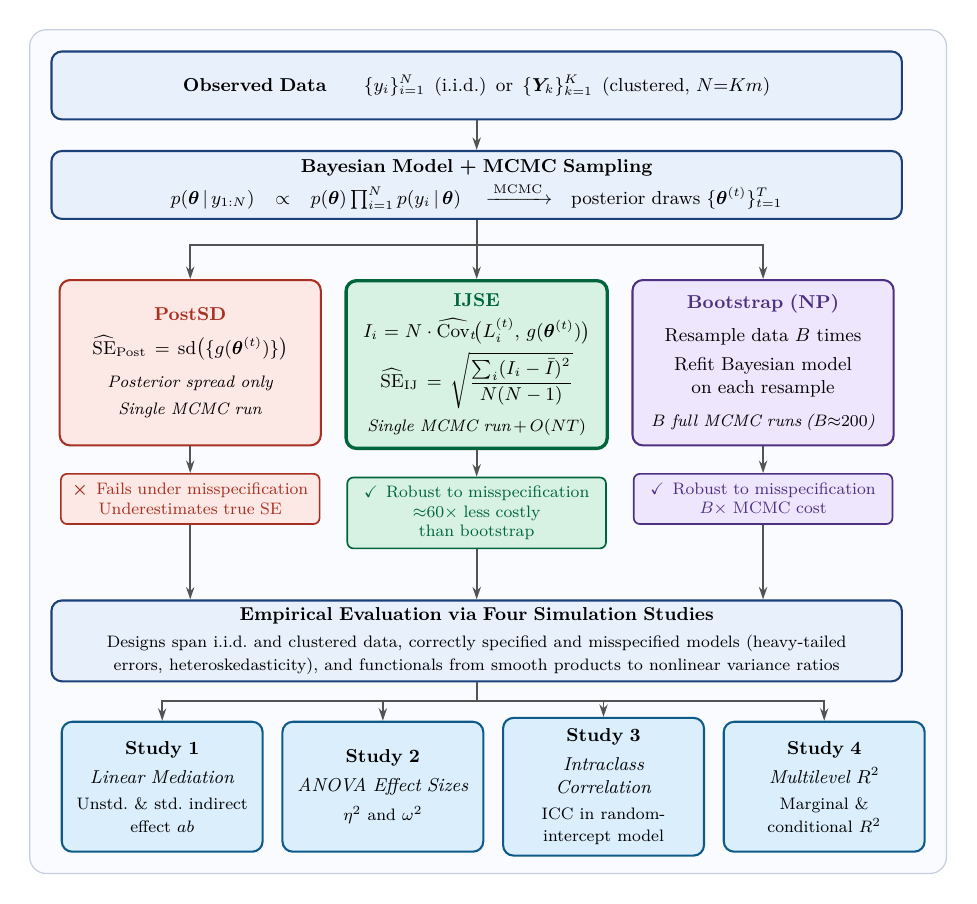}
  \caption{\textbf{Overview of the IJSE methodology and evaluation design.} From a single Bayesian MCMC run, PostSD, IJSE, and the nonparametric bootstrap each estimate the frequentist standard error of a posterior functional; under misspecification PostSD fails while IJSE matches the bootstrap at approximately $60\times$ lower cost. The four simulation studies assess this framework across representative functionals spanning mediation analysis, ANOVA effect sizes, and multilevel modeling.}
  \label{fig:workflow}
\end{figure}

The next four sections evaluate IJSE empirically via simulation studies, each targeting a distinct class of posterior functional that is common in applied research in social and behavioral sciences: a product of regression coefficients (standardized and unstandardized indirect effects in mediation), a ratio of variance components (ANOVA effect sizes $\eta^2$ and $\omega^2$), an intraclass correlation from a random-effects model, and the proportion of variance explained ($R^2$) in a multilevel regression. Together, these cases span i.i.d.\ and clustered data, correctly specified and misspecified models, and functionals ranging from smooth linear contrasts to nonlinear variance ratios, thereby demonstrating that IJSE generalizes well beyond the idealized settings in which influence-function theory is most easily derived. Figure~\ref{fig:workflow} summarizes the complete methodology and evaluation design.

\section{Simulation Study 1: Linear Mediation}
\label{sec:sim1}

Mediation analysis is widely used in the social and behavioral sciences to decompose the total effect of a predictor $X$ on an outcome $Y$ into a direct effect and an indirect effect transmitted through a mediator $M$ \citep{BaronKenny1986,MacKinnon2008}. The indirect effect, typically quantified as the product $ab$ of the mediator regression slope $a$ and the outcome regression slope $b$, captures the mechanism of interest in many substantive theories. Because $ab$ is a nonlinear function of regression coefficients, its sampling distribution is skewed in small to moderate samples \citep{MacKinnon2004}, and standard error estimation has received considerable attention \citep{Sobel1982,ShroutBolger2002}.

Standardized indirect effects serve as scale-free effect-size summaries that improve interpretability and comparability across studies with different measurement scales \citep{PreacherKelley2011,LachowiczPreacherKelley2018}. The standardized slope for a regression of $Y$ on $X$ is
\begin{equation}
\beta_{X\to Y}^{\mathrm{std}}
\;\equiv\;
\beta_{X\to Y}\,\frac{\mathrm{sd}(X)}{\mathrm{sd}(Y)},
\label{eq:std_slope_def}
\end{equation}
interpreted as the expected change in $Y$ (in SD units) per one-SD increase in $X$ \citep{Schielzeth2010,Gelman2008}. However, the standardization in \eqref{eq:std_slope_def} depends on estimated SDs, which are nonlinear functions of second moments sensitive to heavy tails, outliers, and heteroskedasticity \citep{Micceri1989}. Under such deviations, SEs for standardized indirect effects can be severely distorted \citep{YuanMackinnon2014}. The nonparametric bootstrap is often recommended as a robust alternative \citep{BollenStine1990}, but pairing bootstrap resampling with Bayesian posterior computation is costly.

This section evaluates PostSD, IJSE, and the nonparametric bootstrap for two functionals computed from the same mediation model: the unstandardized indirect effect $g_1(\bm{\theta}) = ab$ and the standardized indirect effect $g_2(\bm{\theta}) = ab/\mathrm{sd}(Y)$. Because both functionals share the same posterior draws, they can be evaluated within a single MCMC run, providing a direct comparison of how each method handles a product of location parameters versus a ratio involving variance components.

\subsection{Data-Generating Processes}

We consider a mediation model where an independent variable $X$ influences a dependent variable $Y$ through a mediator $M$. Data for $N$ independent individuals were generated under two conditions.

Under the correctly specified DGP, all variables follow a Gaussian linear model:
\begin{align}
    X_i &\sim \mathcal{N}(0, 1), \label{eq:dgp-x-normal}\\
    M_i &= \alpha_0 + a X_i + \epsilon_{Mi}, \quad \epsilon_{Mi} \sim \mathcal{N}(0, \sigma_M^2), \label{eq:dgp-m-normal}\\
    Y_i &= \beta_0 + c' X_i + b M_i + \epsilon_{Yi}, \quad \epsilon_{Yi} \sim \mathcal{N}(0, \sigma_Y^2). \label{eq:dgp-y-normal}
\end{align}
We set $a=0.3$, $b=0.3$, $c'=0$, $\alpha_0=\beta_0=0$, and $\sigma_M^2=\sigma_Y^2=1$.

The misspecified DGP introduces heteroskedastic and heavy-tailed errors:
\begin{align}
    X_i &\sim \mathcal{N}(0, 1), \label{eq:dgp-x-misspec}\\
    M_i &= \alpha_0 + a X_i + \epsilon_{Mi}, \quad \epsilon_{Mi} \sim \mathrm{Laplace}\left(0, s_M(X_i)\right), \quad s_M(x) = \frac{e^{\kappa_M|x|}}{\sqrt{2}}, \label{eq:dgp-m-misspec} \\
    Y_i &= \beta_0 + c' X_i + b M_i + \epsilon_{Yi}, \quad \epsilon_{Yi} = s_Y(M_i)\cdot t_{\nu}^*, \quad s_Y(m) = e^{\kappa_Y|m|}, \label{eq:dgp-y-misspec}
\end{align}
where $t_{\nu}^* = t_\nu\sqrt{(\nu-2)/\nu}$ is a standardized $t$ random variable with unit variance. To make the standardized functional more sensitive to variance misspecification, we use larger path coefficients than in the correctly specified DGP:
\begin{equation}
a=0.50,\quad b=0.50,\quad c'=0,\quad \nu=3,\quad \kappa_M=0.45,\quad \kappa_Y=0.30.
\label{eq:dgp_params_sim1}
\end{equation}
This construction produces heavy-tailed $Y \mid M$, heteroskedasticity in both stages, and strong dependence of the standardized effect's denominator on variance components.

The simulation crossed the two DGPs with sample sizes $N \in \{200, 500, 1000\}$. For each of the six conditions, $R = 400$ independent datasets were generated. All estimation procedures use Gaussian linear working models for $M_i \mid X_i$ and $Y_i \mid (X_i, M_i)$, even under \eqref{eq:dgp-m-misspec}--\eqref{eq:dgp-y-misspec}.

\subsection{Bayesian Fitting via Gibbs Sampling}

We fit two conjugate Bayesian linear regressions \citep{GelmanBDA3}, one for each stage of the mediation model:
\begin{enumerate}
  \item \emph{Mediator model} ($M_i \mid X_i$): response $y_M = (M_1,\dots,M_N)^\top$, design matrix $X_M = [\mathbf{1},\, \mathbf{x}] \in \mathbb{R}^{N\times 2}$, coefficient vector $\beta_M = (\alpha_0,\, a)^\top$.
  \item \emph{Outcome model} ($Y_i \mid X_i, M_i$): response $y_Y = (Y_1,\dots,Y_N)^\top$, design matrix $X_Y = [\mathbf{1},\, \mathbf{x},\, \mathbf{m}] \in \mathbb{R}^{N\times 3}$, coefficient vector $\beta_Y = (\beta_0,\, c',\, b)^\top$.
\end{enumerate}
For each regression with response $y \in \mathbb{R}^N$ and design matrix $X \in \mathbb{R}^{N \times p}$, we adopt the prior
\begin{equation}
\beta \mid \sigma^2 \sim \mathcal{N}\!\left(0, \tau^2 \sigma^2 I_p\right),
\qquad
\sigma^2 \sim \mathrm{IG}(a_0,b_0),
\label{eq:prior}
\end{equation}
with weakly informative hyperparameters $\tau^2=10^6$ and $a_0=b_0=10^{-2}$. The full conditional posteriors are
\begin{align}
\sigma^2 \mid \beta,y &\sim \mathrm{IG}\!\left(a_0+\tfrac{N+p}{2},\; b_0 + \tfrac12\|y-X\beta\|_2^2 + \tfrac{1}{2\tau^2}\|\beta\|_2^2\right), \label{eq:sigma_cond}\\
\beta \mid \sigma^2,y &\sim \mathcal{N}\!\left(m_n,\; \sigma^2 V_n\right), \label{eq:beta_cond}
\end{align}
where $V_n = (X^\top X + \tau^{-2} I_p)^{-1}$ and $m_n = V_n X^\top y$.

\begin{algorithm}[!htb]
\DontPrintSemicolon
\caption{Gibbs sampler for Bayesian linear regression}
\KwIn{$y\in\mathbb{R}^N$, $X\in\mathbb{R}^{N\times p}$, $T$, $T_{\mathrm{burn}}$, $\tau^2$, $a_0$, $b_0$}
\KwOut{Draws $(\beta^{(t)},\sigma^{2,(t)})_{t=1}^{T}$}
Compute $V_n\leftarrow (X^\top X+\tau^{-2}I)^{-1}$, $m_n\leftarrow V_n X^\top y$\;
Initialize $\beta^{(0)}$ via ridge regression, $\sigma^{2,(0)} \leftarrow \|y-X\beta^{(0)}\|_2^2/(N-p)$\;
\For{$t=1$ \KwTo $T+T_{\mathrm{burn}}$}{
Draw $\sigma^{2,(t)}\sim \mathrm{IG}\left(a_0+\tfrac{N+p}{2}, b_0 + \tfrac12\|y-X\beta^{(t-1)}\|_2^2 + \tfrac{1}{2\tau^2}\|\beta^{(t-1)}\|_2^2\right)$\;
Draw $\beta^{(t)}\sim \mathcal{N}\left(m_n, \sigma^{2,(t)} V_n\right)$\;
\If{$t>T_{\mathrm{burn}}$}{store $(\beta^{(t)},\sigma^{(t)})$}
}
\Return stored draws\;
\end{algorithm}

Running Algorithm~1 on each regression yields posterior draws $\beta_M^{(t)} = (\alpha_0^{(t)}, a^{(t)})^\top$, $\sigma_M^{2,(t)}$, $\beta_Y^{(t)} = (\beta_0^{(t)}, c'^{(t)}, b^{(t)})^\top$, and $\sigma_Y^{2,(t)}$ for $t = 1, \ldots, T$, with $T = 3{,}000$ after a burn-in of $T_{\mathrm{burn}} = 500$.

From a single set of posterior draws, we compute two target functionals. The first is the unstandardized indirect effect,
\begin{equation}
g_1^{(t)} = a^{(t)} b^{(t)},
\label{eq:g1_ie}
\end{equation}
whose posterior mean $\bar{g}_1 = T^{-1}\sum_t g_1^{(t)}$ is the Bayesian point estimator of $ab$.

The second functional is the standardized indirect effect. Because $\mathrm{Var}(X)=1$ by design, the standardized path coefficients yield
\begin{equation}
g_2(\bm{\theta})
=
a_{\mathrm{std}}\,b_{\mathrm{std}}
=
\frac{ab}{\mathrm{sd}(Y)},
\label{eq:g2_std_ie_def}
\end{equation}
where $\mathrm{sd}(Y)$ is the model-implied marginal standard deviation. Under the working model with $\mathrm{Var}(X)=1$,
\begin{equation}
\mathrm{Var}(Y) = (c'+ab)^2 + b^2\sigma_M^2 + \sigma_Y^2,
\label{eq:varY_work}
\end{equation}
so each posterior draw of the standardized indirect effect is
\begin{equation}
g_2^{(t)}
=
\frac{a^{(t)} b^{(t)}}{\sqrt{(c'^{(t)}+a^{(t)}b^{(t)})^2 + b^{(t),2}\,\sigma_M^{2,(t)} + \sigma_Y^{2,(t)}}}.
\label{eq:g2_draw}
\end{equation}
We define the standardized indirect effect using the model-implied marginal SD from \eqref{eq:varY_work}, rather than the sample statistic $s_Y = \mathrm{sd}(\{y_i\}_{i=1}^N)$. This aligns with structural equation modeling conventions, where standardized solutions scale by model-implied variances \citep{Rosseel2012,PreacherKelley2011,PesiganCheung2020}. Treating $g_2(\bm{\theta})$ as a smooth functional of the full parameter vector ensures that posterior uncertainty propagates correctly through the variance components.

\subsection{Methods Compared and Performance Measures}
\label{subsec:sim1_design}

The posterior SD is obtained directly from the draws $\{g_j^{(t)}\}_{t=1}^T$ for $j = 1, 2$ as $\widehat{\mathrm{SE}}_{\mathrm{PostSD}} = \mathrm{sd}(\{g_j^{(t)}\})$. For the nonparametric bootstrap, we resample rows with replacement for $b = 1, \ldots, B$, refit both models via Gibbs, compute the posterior means $\bar{g}_{1,b}^*$ and $\bar{g}_{2,b}^*$, and report $\widehat{\mathrm{SE}}_{\mathrm{NP}} = \mathrm{sd}(\{\bar{g}_{j,b}^*\})$. We use $B = 100$ bootstrap resamples, each with $T_{\mathrm{boot}} = 800$ iterations and a burn-in of 200. For the infinitesimal jackknife, we use the baseline MCMC draws to compute the per-observation log-likelihood
\begin{equation}
    L_i^{(t)} = \log\phi\big(m_i;\,\mu_{M,i}^{(t)},\sigma_M^{(t)}\big) + \log\phi\big(y_i;\,\mu_{Y,i}^{(t)},\sigma_Y^{(t)}\big),
    \label{eq:Li_mediation}
\end{equation}
where $\phi(\cdot;\mu,\sigma)$ denotes the $\mathcal{N}(\mu,\sigma^2)$ density. The log-likelihood matrix is computed once and reused for both functionals. For each $g_j$ ($j = 1, 2$), compute $I_i$ via \eqref{eq:Ii_formula} and $\widehat{\mathrm{SE}}_{\mathrm{IJ}}$ via \eqref{eq:ijse_formula}.

We evaluate each method using the following performance measures. Let $\bar{g}_{j,r}$ denote the posterior mean for functional $g_j$ on replication $r = 1, \ldots, R$. The Monte Carlo benchmark for the frequentist SE is
\begin{equation}
\mathrm{SE}_{\mathrm{MC}}
\;\equiv\;
\mathrm{sd}\!\left(\{\bar{g}_{j,r}\}_{r=1}^R\right).
\label{eq:se_mc_sim1}
\end{equation}
For each method $m \in \{\mathrm{PostSD}, \mathrm{NP}, \mathrm{IJ}\}$, we report the following measures \citep{MorrisEtAl2019}:
\begin{enumerate}
  \item \emph{Mean SE} ($\overline{\mathrm{SE}}_m$): average of the SE estimates across replications;
  \item \emph{Bias}: $\mathrm{Bias}_m = \overline{\mathrm{SE}}_m - \mathrm{SE}_{\mathrm{MC}}$;
  \item \emph{Relative error}: $\mathrm{RelErr}_m = \mathrm{Bias}_m / \mathrm{SE}_{\mathrm{MC}}$;
  \item \emph{Empirical coverage} (EC): empirical proportion of replications for which $\bar{g}_{j,r} \pm 1.96\,\widehat{\mathrm{SE}}_r$ contains the grand mean $\bar{\bar{g}}_j = R^{-1}\sum_r \bar{g}_{j,r}$;
  \item \emph{Mean runtime}: average wall-clock seconds per replication.
\end{enumerate}

\subsection{Results}

\begin{figure}[!htb]
   \centering
   \includegraphics[width=.95\linewidth]{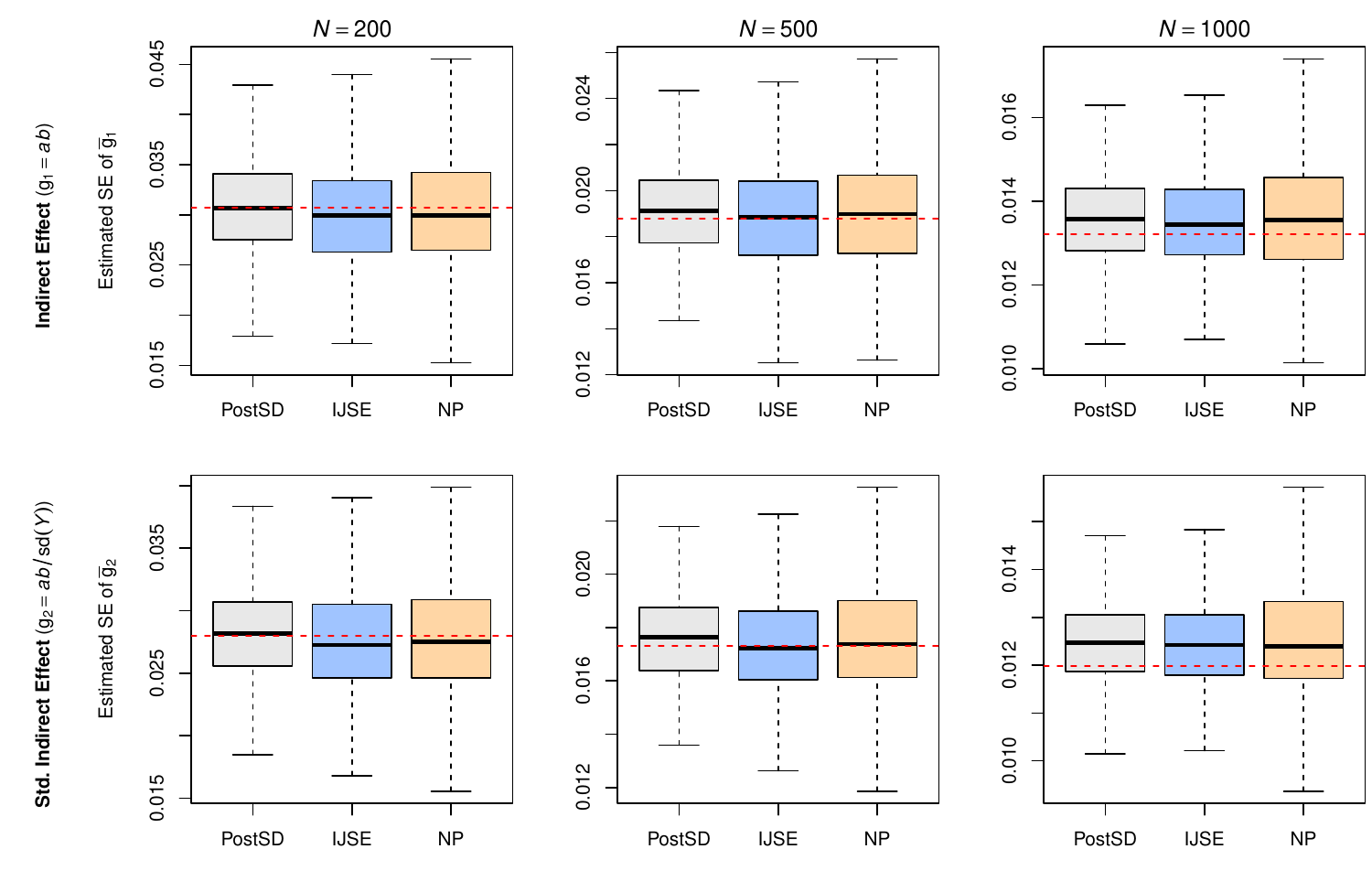}
   \caption{SE estimates under correct specification ($R=400$). Top row: indirect effect $g_1 = ab$; bottom row: standardized indirect effect $g_2 = ab/\mathrm{sd}(Y)$. Dashed red lines indicate the Monte Carlo SE benchmark. All three methods produce similar distributions across sample sizes for both functionals.}
   \label{fig:med-se-correct}
\end{figure}

\begin{figure}[!htb]
   \centering
   \includegraphics[width=.95\linewidth]{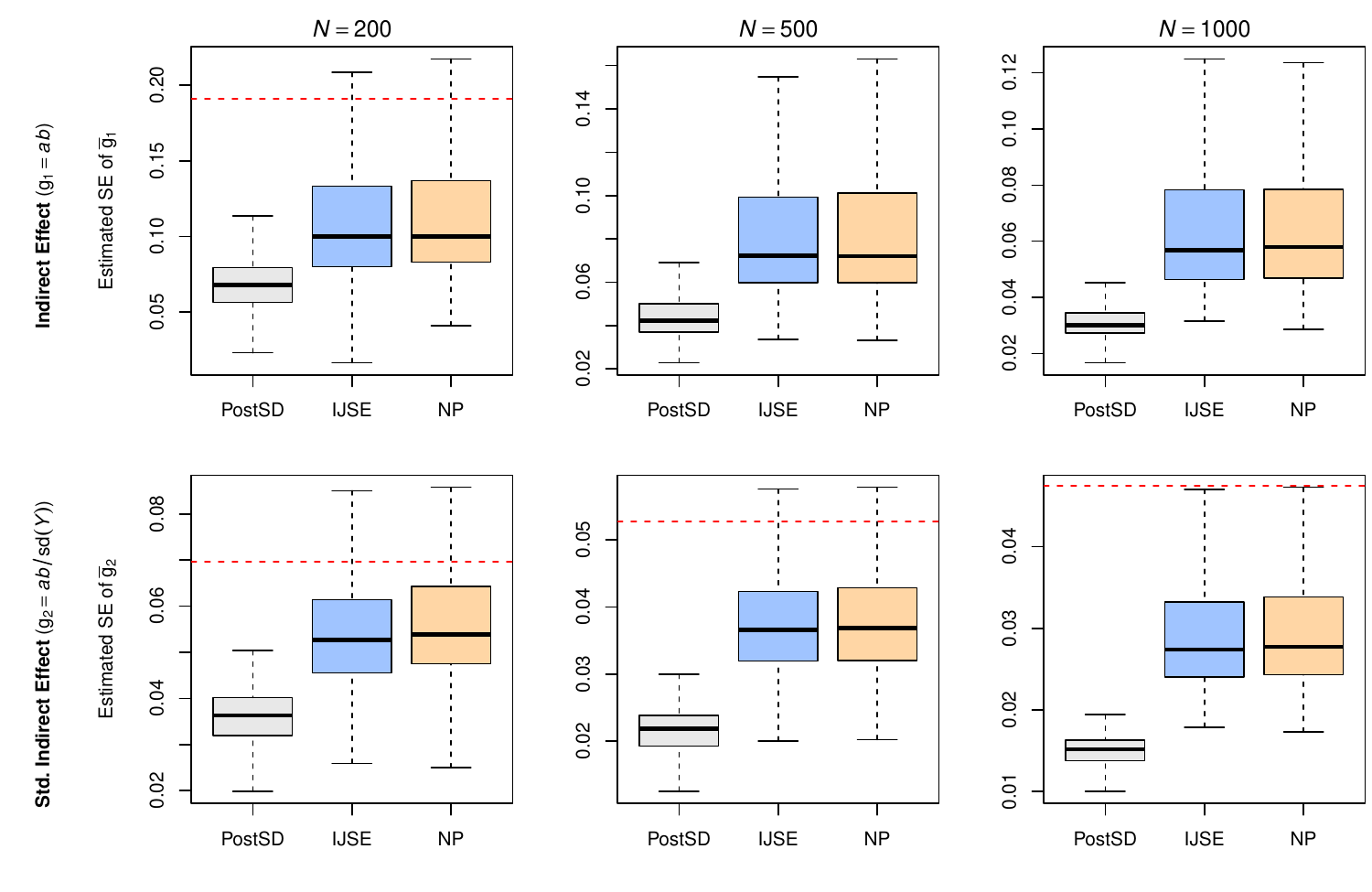}
   \caption{SE estimates under misspecification ($R=400$). Top row: $g_1 = ab$; bottom row: $g_2 = ab/\mathrm{sd}(Y)$. Dashed red lines indicate the Monte Carlo SE. PostSD severely underestimates variability; IJSE and NP agree closely and track the benchmark much more accurately.}
   \label{fig:med-se-misspec}
\end{figure}

\begin{figure}[!htb]
   \centering
   \includegraphics[width=.85\linewidth]{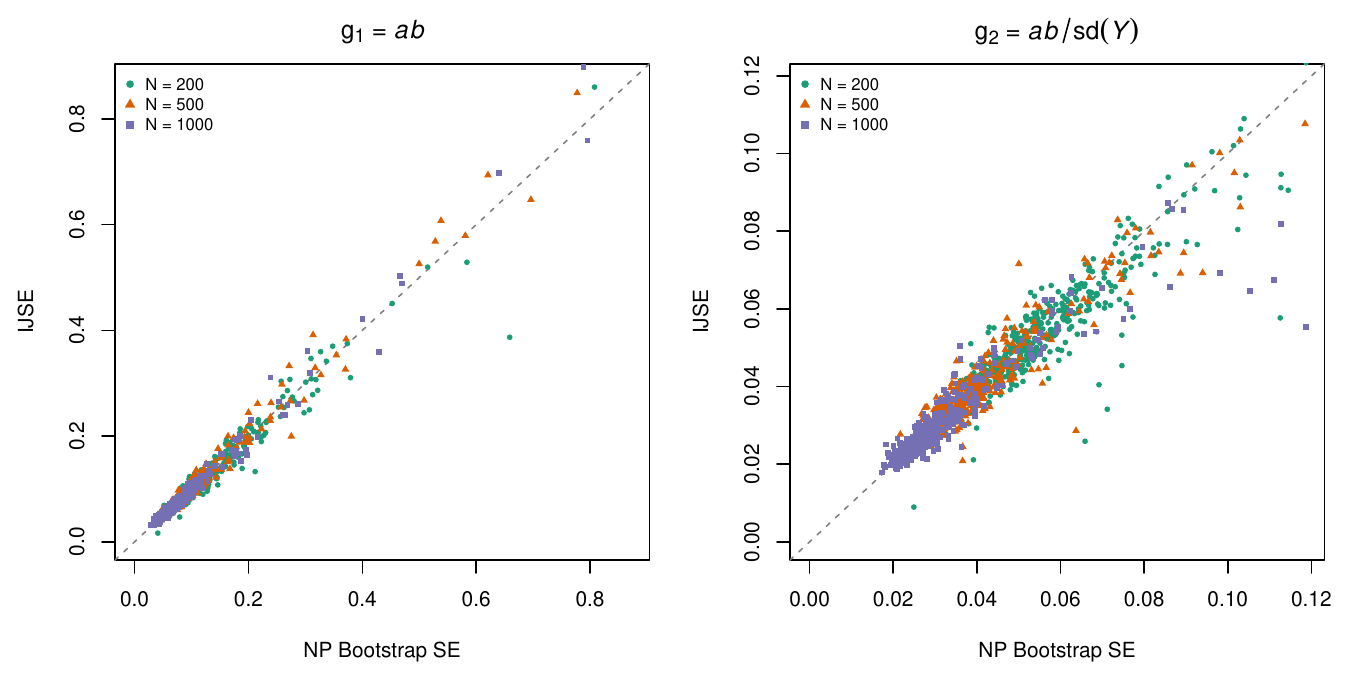}
   \caption{IJSE vs.\ nonparametric bootstrap SE under misspecification, for the indirect effect (left) and the standardized indirect effect (right). Points cluster tightly around $y=x$, indicating close replication-level agreement.}
   \label{fig:med-scatter}
\end{figure}

\begin{figure}[!htb]
   \centering
   \includegraphics[width=.60\linewidth]{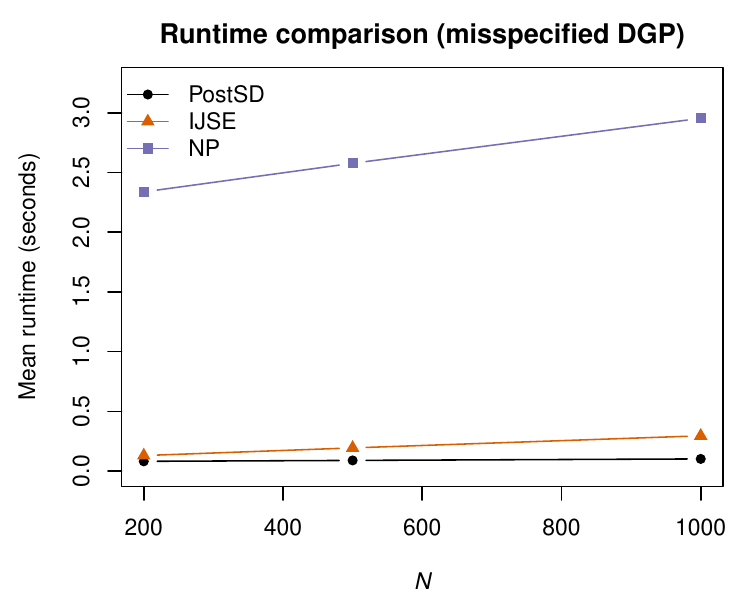}
   \caption{Mean runtime by method and sample size (misspecified DGP). IJSE adds modest overhead to PostSD while remaining an order of magnitude faster than the nonparametric bootstrap.}
   \label{fig:med-rt}
\end{figure}

Figures~\ref{fig:med-se-correct}--\ref{fig:med-rt} display the distributions of SE estimates and runtimes across the 400 replications. Table~\ref{tab:mediation_results} then provides the full numerical summaries for both functionals under correct specification and misspecification.

Under correct specification, the three box plots in each panel of Figure~\ref{fig:med-se-correct} are nearly indistinguishable, and all three medians sit close to the Monte Carlo benchmark (dashed red line). The distributions shrink as $N$ increases from 200 to 1000, as expected. No systematic separation appears between PostSD, IJSE, and the nonparametric bootstrap for either functional, confirming that all three methods are well calibrated when the Gaussian working model matches the true data-generating process.

Figure~\ref{fig:med-se-misspec} reveals a strikingly different picture under misspecification. The PostSD box is shifted well below the Monte Carlo benchmark across all sample sizes and both functionals, reflecting the severe underestimation caused by the Gaussian working likelihood's inability to detect the excess variability from heteroskedastic $t_3$ errors. IJSE and the nonparametric bootstrap, by contrast, produce box plots that overlap substantially and sit much closer to the benchmark. The gap between PostSD and the other two methods widens as $N$ grows, because the Gaussian posterior concentrates around incorrect variance parameters at a rate that outpaces its ability to capture the true sampling variability.

The replication-level agreement between IJSE and the bootstrap is further confirmed in Figure~\ref{fig:med-scatter}, where points for both functionals cluster tightly along the identity line. Correlations between the two SE estimates exceed 0.93 for $g_1$ and 0.90 for $g_2$ across all sample sizes under misspecification, indicating that the infinitesimal jackknife captures essentially the same information about observation-level influence as the full resampling procedure.

Figure~\ref{fig:med-rt} displays the computational cost. PostSD required 0.08--0.11\,s per replication, IJSE added modest overhead at 0.13--0.41\,s, and the nonparametric bootstrap was 10 to 23 times slower at 2.3--3.0\,s per replication owing to the $B=100$ refits. The IJSE overhead is dominated by the $O(NT)$ log-likelihood covariance computation and scales linearly with sample size.

\begin{table}[!htb]
\centering
\caption{Simulation results for the indirect effect $g_1 = ab$ and the standardized indirect effect $g_2 = ab/\mathrm{sd}(Y)$ ($R=400$ replications per condition). $\mathrm{SE}_{\mathrm{MC}}$: Monte Carlo benchmark; Bias: Mean SE $-$ $\mathrm{SE}_{\mathrm{MC}}$; RelErr: Bias$/\mathrm{SE}_{\mathrm{MC}}$; EC: empirical coverage of 95\% intervals.}
\label{tab:mediation_results}
\small
\setlength{\tabcolsep}{3.5pt}
\begin{tabular}{cl cccrc cccrc r}
\toprule
& & \multicolumn{5}{c}{Indirect Effect ($g_1 = ab$)}
  & \multicolumn{5}{c}{Std.\ Indirect Effect ($g_2 = ab/\mathrm{sd}(Y)$)}
  & \\
\cmidrule(lr){3-7} \cmidrule(lr){8-12}
$N$ & Method
  & $\mathrm{SE}_{\mathrm{MC}}$ & Mean SE & Bias & RelErr & EC
  & $\mathrm{SE}_{\mathrm{MC}}$ & Mean SE & Bias & RelErr & EC
  & Time (s) \\
\midrule
\multicolumn{13}{l}{\textit{Panel A: Correct specification}} \\[2pt]
200 & PostSD & 0.031 & 0.031 & 0.000 & $+1\%$ & 0.94
             & 0.028 & 0.028 & 0.000 & $+1\%$ & 0.94 & 0.09 \\
    & IJSE   &       & 0.030 & $-0.001$ & $-1\%$ & 0.93
             &       & 0.028 & 0.000 & $-1\%$ & 0.93 & 0.17 \\
    & NP     &       & 0.031 & 0.000 & $+0\%$ & 0.93
             &       & 0.028 & 0.000 & $+0\%$ & 0.94 & 2.71 \\[2pt]
500 & PostSD & 0.019 & 0.019 & 0.000 & $+2\%$ & 0.94
             & 0.017 & 0.018 & 0.000 & $+2\%$ & 0.94 & 0.10 \\
    & IJSE   &       & 0.019 & 0.000 & $+1\%$ & 0.94
             &       & 0.017 & 0.000 & $+0\%$ & 0.94 & 0.26 \\
    & NP     &       & 0.019 & 0.000 & $+2\%$ & 0.94
             &       & 0.018 & 0.000 & $+1\%$ & 0.94 & 2.82 \\[2pt]
1000 & PostSD & 0.013 & 0.014 & 0.000 & $+3\%$ & 0.96
              & 0.012 & 0.012 & 0.000 & $+4\%$ & 0.96 & 0.11 \\
     & IJSE   &       & 0.014 & 0.000 & $+2\%$ & 0.96
              &       & 0.012 & 0.000 & $+4\%$ & 0.97 & 0.41 \\
     & NP     &       & 0.014 & 0.000 & $+3\%$ & 0.95
              &       & 0.013 & 0.001 & $+4\%$ & 0.96 & 2.99 \\
\midrule
\multicolumn{13}{l}{\textit{Panel B: Misspecified DGP}} \\[2pt]
200 & PostSD & 0.191 & 0.073 & $-0.118$ & $-62\%$ & 0.71
             & 0.070 & 0.035 & $-0.034$ & $-49\%$ & 0.73 & 0.08 \\
    & IJSE   &       & 0.130 & $-0.061$ & $-32\%$ & 0.93
             &       & 0.056 & $-0.014$ & $-19\%$ & 0.92 & 0.13 \\
    & NP     &       & 0.133 & $-0.058$ & $-30\%$ & 0.94
             &       & 0.059 & $-0.011$ & $-16\%$ & 0.94 & 2.34 \\[2pt]
500 & PostSD & 0.179 & 0.047 & $-0.132$ & $-74\%$ & 0.59
             & 0.053 & 0.021 & $-0.031$ & $-59\%$ & 0.60 & 0.09 \\
    & IJSE   &       & 0.106 & $-0.073$ & $-41\%$ & 0.92
             &       & 0.040 & $-0.012$ & $-24\%$ & 0.88 & 0.19 \\
    & NP     &       & 0.105 & $-0.074$ & $-41\%$ & 0.92
             &       & 0.041 & $-0.012$ & $-22\%$ & 0.90 & 2.58 \\[2pt]
1000 & PostSD & 0.201 & 0.034 & $-0.167$ & $-83\%$ & 0.58
              & 0.047 & 0.015 & $-0.033$ & $-69\%$ & 0.57 & 0.10 \\
     & IJSE   &       & 0.098 & $-0.104$ & $-51\%$ & 0.94
              &       & 0.031 & $-0.016$ & $-34\%$ & 0.89 & 0.30 \\
     & NP     &       & 0.097 & $-0.104$ & $-52\%$ & 0.93
              &       & 0.033 & $-0.015$ & $-31\%$ & 0.92 & 2.95 \\
\bottomrule
\end{tabular}
\end{table}

Table~\ref{tab:mediation_results} quantifies the patterns visible in the figures. Under correct specification (Panel~A), all three methods yield relative errors within $\pm 4\%$ and coverage rates between 93\% and 97\% for both $g_1$ and $g_2$, close to the nominal 95\% level. Biases are negligible across all sample sizes, confirming that PostSD, IJSE, and the nonparametric bootstrap are all valid when the Gaussian working model matches the true DGP. The two functionals behave similarly because the working model captures both the location and scale structure of the data under correct specification.

Under misspecification (Panel~B), PostSD exhibits severe underestimation, with relative errors ranging from $-62\%$ to $-83\%$ for $g_1$ and from $-49\%$ to $-69\%$ for $g_2$. Coverage collapses to 58--71\% for $g_1$ and 57--73\% for $g_2$, far below the nominal level. IJSE and the nonparametric bootstrap perform substantially better, with relative errors that differ by at most 3~percentage points from each other in every condition. For the unstandardized indirect effect $g_1$, both methods achieve coverage of 92--94\%, while for the standardized indirect effect $g_2$, IJSE reaches 88--92\% and the bootstrap 90--94\%.

Both IJSE and the bootstrap exhibit some residual downward bias relative to the Monte Carlo benchmark under misspecification. This arises because the $t_3$ distribution with exponential heteroskedasticity generates occasional extreme observations with disproportionate influence. These outlier-driven replications inflate $\mathrm{SE}_{\mathrm{MC}}$ substantially, creating a benchmark that any fixed-sample SE estimator will tend to underestimate. Despite this, IJSE and the bootstrap maintain coverage rates far above PostSD, because they correctly detect the heightened variability in the data at hand through their sensitivity to individual observation influence.

The standardized functional $g_2$ is more forgiving than $g_1$ under misspecification. PostSD's relative error for $g_2$ ($-49\%$ to $-69\%$) is less extreme than for $g_1$ ($-62\%$ to $-83\%$). This occurs because the denominator $\mathrm{sd}(Y)$ partially absorbs the excess variability caused by heavy tails, attenuating the scale of $g_2$ in replications where outliers inflate both numerator and denominator. The same mechanism yields smaller Monte Carlo SEs for $g_2$ (0.047--0.070) than for $g_1$ (0.179--0.201), making the relative estimation task more tractable.

In summary, IJSE closely approximates the bootstrap SE at a fraction of the computational cost for both the unstandardized and standardized indirect effects, and both methods dominate PostSD under misspecification in terms of bias, coverage, and relative error. The results demonstrate that a single MCMC run suffices to produce calibrated frequentist SEs for multiple functionals simultaneously, including those that involve variance components.

\section{Simulation Study 2: ANOVA Effect Sizes}
\label{sec:sim3}

In psychological research, ANOVA-style effect sizes, $\eta^2$, partial $\eta^2$, generalized $\eta^2$, and $\omega^2$, are routinely reported to quantify experimental effects and to support meta-analysis \citep{Lakens2013,LevineHullett2002,OlejnikAlgina2003,KroesFinley2023}. These measures are nonlinear ratios of location and scale parameters whose sampling distributions can be skewed, especially in small samples \citep{Okada2013,KroesFinley2023}.
Real behavioral outcomes frequently deviate from homoskedastic Gaussian assumptions \citep{Micceri1989,Wilcox2012}, yet analysts commonly fit Gaussian working models. Under such misspecification, PostSD can fail to capture the repeated-sampling variability of $\bar{g}$ (Section~\ref{sec:sandwich}), motivating the use of IJSE \citep{GiordanoBroderick2023}.

\subsection{Data-Generating Process}

We consider a one-way between-subjects design with $J=5$ experimental groups and equal cell sizes $n=N/J$. Let $G_i\in\{1,\dots,J\}$ be the group indicator with exactly $n$ observations per group. Outcomes are generated via
\begin{align}
Y_i &= \mu_{G_i} + \varepsilon_i, \qquad i=1,\dots,N, \label{eq:anova_dgp_y}\\
\mu_j &= \delta\cdot\left(j-\frac{J+1}{2}\right), \qquad j=1,\dots,J, \label{eq:anova_dgp_means}\\
\varepsilon_i \mid (G_i=j) &\equiv s_j \cdot t_{3,i},
\qquad
s_j \equiv \exp(\gamma|\mu_j|)\sqrt{\frac{3-2}{3}}, \label{eq:anova_dgp_err}
\end{align}
where $t_{3,i}$ are i.i.d.\ standard Student-$t$ with $\nu=3$ degrees of freedom. The factor $\sqrt{(3-2)/3}$ rescales $t_3$ to unit variance before the heteroskedastic multiplier $\exp(\gamma|\mu_j|)$. We fix $(\delta,\gamma)=(0.4,0.35)$ to induce a moderate mean separation and pronounced group-dependent heteroskedasticity. This construction mimics a common failure mode: groups with more extreme means also show larger dispersion and heavier tails, which is empirically plausible in behavioral outcomes \citep{Micceri1989,Wilcox2012}.

The simulation varies the total sample size $N\in\{200,400,600\}$, implying $n\in\{40,80,120\}$ per group. For each sample size, $R=300$ independent datasets are generated from \eqref{eq:anova_dgp_y}--\eqref{eq:anova_dgp_err}.

\subsection{Bayesian Fitting and Methods Compared}

The working model is a homoskedastic Gaussian linear regression with group indicators:
\begin{equation}
Y_i \mid \bm{\beta},\sigma^2 \sim \mathcal{N}(x_i^\top \bm{\beta},\,\sigma^2),
\qquad
x_i = (1,\mathbbm{1}\{ G_i=2 \},\dots,\mathbbm{1}\{ G_i=J \})^\top.
\label{eq:anova_working}
\end{equation}
Let $\hat\mu_j(\bm{\beta})$ denote the model-implied mean for group $j$ (grand intercept plus the corresponding indicator coefficient). The target functional is the model-based proportion of variance explained by group membership,
\begin{equation}
g_3(\bm{\theta})
\equiv
\eta^2(\bm{\theta})
=
\frac{\mathrm{Var}_{j}\!\big(\hat\mu_j(\bm{\beta})\big)}
{\mathrm{Var}_{j}\!\big(\hat\mu_j(\bm{\beta})\big)+\sigma^2},
\label{eq:anova_eta_param}
\end{equation}
where $\mathrm{Var}_j(\cdot)$ is the variance across the $J$ group means under equal weights. This is a smooth ratio of regression parameters and the residual variance. Under misspecification, $g_3(\bm{\theta})$ is interpreted as the effect size induced by the pseudo-true Gaussian approximation \citep{White1982,KleijnVanDerVaart2012,Muller2013}.

We adopt the same conjugate normal--inverse-gamma prior as in Section~\ref{sec:sim1},
\begin{equation}
\bm{\beta}\mid\sigma^2 \sim \mathcal{N}\!\left(\bm{0},\,\tau^2\sigma^2 I_p\right),
\qquad
\sigma^2 \sim \mathrm{IG}(a_0,b_0),
\label{eq:anova_prior}
\end{equation}
with $\tau^2=10^6$ and $a_0=b_0=10^{-2}$. Conditional conjugacy yields standard two-block Gibbs updates for $(\bm{\beta},\sigma^2)$, where the $\bm{\beta}$-update is a single multivariate normal draw with precomputable sufficient statistics $X^\top X$ and $X^\top y$.

PostSD and the nonparametric bootstrap are computed as described in Section~\ref{subsec:sim1_design}, with $B=100$ bootstrap resamples. For IJSE, the per-observation log-likelihood under the working model is
\begin{equation}
L_i^{(t)}=\log\phi\!\left(Y_i;\,x_i^\top\bm{\beta}^{(t)},\,\sigma^{(t)}\right),
\qquad i=1,\dots,N,
\label{eq:anova_li}
\end{equation}
and the influence proxies $I_i$ and the IJ variance estimator follow \eqref{eq:Ii_formula}--\eqref{eq:ijse_formula}. Performance measures are the same as in Section~\ref{subsec:sim1_design}.

\subsection{Results}
\label{subsec:results_anova}

Table~\ref{tab:anova_results} presents the simulation results for $\eta^2$ under the heteroskedastic heavy-tailed DGP.

\begin{table}[!htb]
\centering
\caption{Simulation results for the ANOVA effect size $g_3(\bm{\theta}) = \eta^2$ under heteroskedastic $t_3$ errors ($R=300$ replications per $N$). Column definitions follow Table~\ref{tab:mediation_results}.}
\label{tab:anova_results}
\small
\begin{tabular}{clcccrcr}
\toprule
$N$ & Method & $\mathrm{SE}_{\mathrm{MC}}$ & Mean SE & Bias & RelErr & EC & Time (s) \\
\midrule
200 & PostSD & 0.059 & 0.046 & $-0.013$ & $-21\%$ & 0.85 & 0.04 \\
    & IJSE   &       & 0.054 & $-0.005$ & $-9\%$  & 0.89 & 0.09 \\
    & NP     &       & 0.057 & $-0.002$ & $-4\%$  & 0.92 & 2.41 \\[2pt]
400 & PostSD & 0.049 & 0.033 & $-0.016$ & $-33\%$ & 0.83 & 0.06 \\
    & IJSE   &       & 0.042 & $-0.007$ & $-15\%$ & 0.90 & 0.15 \\
    & NP     &       & 0.043 & $-0.006$ & $-12\%$ & 0.92 & 2.89 \\[2pt]
600 & PostSD & 0.040 & 0.027 & $-0.013$ & $-33\%$ & 0.84 & 0.06 \\
    & IJSE   &       & 0.036 & $-0.005$ & $-11\%$ & 0.92 & 0.19 \\
    & NP     &       & 0.037 & $-0.004$ & $-9\%$  & 0.93 & 2.86 \\
\bottomrule
\end{tabular}
\end{table}

PostSD underestimates $\mathrm{SE}_{\mathrm{MC}}$ by 21--33\%, with coverage rates of 83--85\%, reflecting the Gaussian working model's failure to account for group-dependent dispersion and heavy tails. The underestimation worsens from $N=200$ to $N=400$ and remains stable at $N=600$, indicating that the misspecification bias does not vanish with increasing sample size.

IJSE and the NP bootstrap agree within 3 to 5 percentage points of relative error across all sample sizes. IJSE achieves coverage of 89--92\%, while NP reaches 92--93\%, both substantially closer to the nominal 95\% than PostSD. In terms of computation, IJSE runs in 0.09--0.19\,s per replication, adding modest overhead to PostSD (0.04--0.06\,s), while the bootstrap requires 2.4--2.9\,s, roughly 15 to 27 times slower than IJSE.


\section{Simulation Study 3: Intraclass Correlation}
\label{sec:sim4}

The intraclass correlation coefficient (ICC) measures the proportion of outcome variance attributable to between-cluster differences and is central to rater reliability, multilevel study design, and computation of design effects \citep{ShroutFleiss1979,McGrawWong1996,RaudenbushBryk2002,RaykovMarcoulides2015}. ICC estimation is sensitive in small to moderate samples and can be affected by deviations from Gaussian random-effects assumptions \citep{Micceri1989,Wilcox2012,TenHoveEtAl2025}. Like the ANOVA effect sizes above, the ICC is a nonlinear ratio of variance components for which PostSD may be miscalibrated under misspecification. This simulation also introduces a fixed-effect covariate into the random-intercept model, which enables the study of multilevel $R^2$ measures from the same fitted model in Section~\ref{sec:sim_r2}.

\subsection{Data-Generating Process}
\label{subsec:icc_dgp}

Consider a random-intercept model with a single covariate. The design comprises $K$ clusters, each containing a fixed number of $m=5$ units, so $N=Km$. For cluster $k=1,\dots,K$ and unit $i=1,\dots,m$:
\begin{align}
x_{ik} &\sim \mathcal{N}(0,1), \quad \text{i.i.d.}, \label{eq:icc_dgp_x}\\
Y_{ik} &= \mu + \beta\, x_{ik} + U_k + \varepsilon_{ik}, \label{eq:icc_dgp_y}\\
U_k &\equiv \sigma_U \cdot \sqrt{\frac{\nu-2}{\nu}} \cdot t_{\nu,k}, \label{eq:icc_dgp_u}\\
\varepsilon_{ik}\mid U_k &\equiv \sigma_\varepsilon \cdot \exp(\lambda|U_k|)\cdot \frac{1}{\sqrt{2}}\,\ell_{ik},
\qquad
\ell_{ik}\sim \mathrm{Laplace}(0,1), \label{eq:icc_dgp_e}
\end{align}
where $t_{\nu,k}$ are i.i.d.\ standard $t_\nu$ and $\ell_{ik}$ are i.i.d.\ standard Laplace with $\mathrm{Var}(\ell_{ik})=2$. The prefactors rescale $t_\nu$ and Laplace to unit variance before applying $\sigma_U$ and $\sigma_\varepsilon$, and the multiplier $\exp(\lambda|U_k|)$ induces heteroskedasticity correlated with the random intercept. The covariate $x_{ik}$ is drawn independently of the cluster structure, so its variance-explained contribution is entirely at the fixed-effect level.

We fix $(\mu,\beta,\sigma_U^2,\sigma_\varepsilon^2,\lambda,\nu)=(0,\;0.5,\;0.30,\;1.70,\;0.25,\;3)$, yielding a moderate clustering signal with frequent outliers and cluster-dependent dispersion. With $\mathrm{Var}(x)\approx 1$, the fixed-effect variance is $\sigma_f^2 = \beta^2\,\mathrm{Var}(x) \approx 0.25$.

The simulation varies $K\in\{40,80,120\}$, corresponding to $N\in\{200,400,600\}$. For each value of $K$, $R=300$ independent datasets are generated from \eqref{eq:icc_dgp_x}--\eqref{eq:icc_dgp_e}.

\subsection{Bayesian Fitting and Methods Compared}
\label{subsec:icc_fitting}

The working model is a Gaussian random-intercept model with a fixed-effect covariate:
\begin{equation}
Y_{ik}\mid \mu,\beta, U_k,\sigma_\varepsilon^2 \sim \mathcal{N}(\mu + \beta\,x_{ik} + U_k,\;\sigma_\varepsilon^2),
\qquad
U_k\mid \sigma_U^2 \sim \mathcal{N}(0,\sigma_U^2).
\label{eq:icc_working}
\end{equation}
The target functional for this section is the ICC,
\begin{equation}
g_4(\bm{\theta})
\equiv
\rho(\bm{\theta})
=
\frac{\sigma_U^2}{\sigma_U^2+\sigma_\varepsilon^2}.
\label{eq:icc_rho}
\end{equation}
Under misspecification, $\rho(\bm{\theta})$ is interpreted as the variance ratio implied by the pseudo-true Gaussian approximation \citep{White1982,KleijnVanDerVaart2012}.

We adopt weakly informative priors:
\begin{equation}
\mu \sim \mathcal{N}(0,\kappa^2),\qquad
\beta \sim \mathcal{N}(0,\tau_\beta^2),\qquad
\sigma_\varepsilon^2 \sim \mathrm{IG}(a_\varepsilon,b_\varepsilon),\qquad
\sigma_U^2 \sim \mathrm{IG}(a_U,b_U),
\label{eq:icc_prior}
\end{equation}
with large $\kappa^2$ and $\tau_\beta^2$ and small $(a_\varepsilon,b_\varepsilon,a_U,b_U)$. All full conditionals are conjugate, yielding a five-block Gibbs sampler. The updates for $U_k$, $\mu$, $\sigma_\varepsilon^2$, and $\sigma_U^2$ follow standard random-intercept conjugacy \citep{GelmanBDA3}, and the additional block for $\beta$ is
\begin{equation}
\beta \mid - \;\sim\; \mathcal{N}\!\left(
\frac{\sum_{k}\sum_i x_{ik}(Y_{ik}-\mu-U_k)/\sigma_\varepsilon^2}
     {\sum_{k}\sum_i x_{ik}^2/\sigma_\varepsilon^2 + 1/\tau_\beta^2},
\;\;
\left(\frac{\sum_{k}\sum_i x_{ik}^2}{\sigma_\varepsilon^2} + \frac{1}{\tau_\beta^2}\right)^{-1}
\right).
\label{eq:icc_beta_update}
\end{equation}
The remaining four blocks take the same form as in a standard random-intercept model, with all residuals computed as $Y_{ik} - \mu - \beta\,x_{ik} - U_k$. For each posterior draw, $\rho^{(t)}$ is computed via \eqref{eq:icc_rho}.

PostSD and the nonparametric bootstrap are computed as described in Section~\ref{subsec:sim1_design}. Because the data are clustered, the bootstrap resamples entire clusters $\{(Y_{ik}, x_{ik})\}_{i=1}^{m}$ with replacement, preserving the within-cluster covariate structure, with $B=100$ resamples. For IJSE, the per-cluster log-likelihood contribution follows \eqref{eq:cluster_Lk}:
\begin{equation}
L_k^{(t)}
=
\log \phi\!\left(U_k^{(t)};0,\sigma_U^{(t)}\right)
+
\sum_{i=1}^m \log \phi\!\left(Y_{ik};\,\mu^{(t)}+\beta^{(t)}x_{ik}+U_k^{(t)},\,\sigma_\varepsilon^{(t)}\right),
\qquad k=1,\dots,K.
\label{eq:icc_Lk}
\end{equation}
The influence proxies $I_k$ and the IJ variance estimator then follow \eqref{eq:cluster_ijse} with Algorithm~\ref{alg:ijse_cluster}. This log-likelihood matrix is computed once and reused for both the ICC functional in this section and the $R^2$ functionals in Section~\ref{sec:sim_r2}. Performance measures are the same as in Section~\ref{subsec:sim1_design}.

\subsection{Results}
\label{subsec:results_icc}

Table~\ref{tab:icc_results} presents the simulation results for the ICC under heavy-tailed random effects and heteroskedastic within-cluster errors.

\begin{table}[!htb]
\centering
\caption{Simulation results for the intraclass correlation $g_4(\bm{\theta}) = \sigma_U^2/(\sigma_U^2+\sigma_\varepsilon^2)$ under heavy-tailed heteroskedastic errors ($R=300$ replications per $K$). Column definitions follow Table~\ref{tab:mediation_results}.}
\label{tab:icc_results}
\small
\begin{tabular}{cclcccrcr}
\toprule
$K$ & $N$ & Method & $\mathrm{SE}_{\mathrm{MC}}$ & Mean SE & Bias & RelErr & EC & Time (s) \\
\midrule
40 & 200 & PostSD & 0.082 & 0.056 & $-0.026$ & $-32\%$ & 0.83 & 0.04 \\
   &     & IJSE   &       & 0.056 & $-0.026$ & $-32\%$ & 0.73 & 0.08 \\
   &     & NP     &       & 0.055 & $-0.027$ & $-33\%$ & 0.79 & 1.14 \\[2pt]
80 & 400 & PostSD & 0.075 & 0.043 & $-0.031$ & $-42\%$ & 0.75 & 0.05 \\
   &     & IJSE   &       & 0.052 & $-0.023$ & $-30\%$ & 0.77 & 0.12 \\
   &     & NP     &       & 0.052 & $-0.023$ & $-31\%$ & 0.80 & 1.54 \\[2pt]
120 & 600 & PostSD & 0.054 & 0.036 & $-0.018$ & $-34\%$ & 0.78 & 0.07 \\
    &     & IJSE   &       & 0.044 & $-0.010$ & $-19\%$ & 0.81 & 0.15 \\
    &     & NP     &       & 0.042 & $-0.012$ & $-22\%$ & 0.83 & 1.88 \\
\bottomrule
\end{tabular}
\end{table}

With only 40 clusters, the effective sample size for the between-cluster variance component is small, and all three methods underestimate $\mathrm{SE}_{\mathrm{MC}}$ by approximately 32--33\%. Coverage ranges from 73\% (IJSE) to 83\% (PostSD). The IJ approximation requires sufficient independent units (here clusters) to stabilize the influence-function variance, and $K=40$ appears too small for reliable performance.

At $K=80$ and $K=120$, IJSE substantially outperforms PostSD. PostSD's relative error reaches $-42\%$ at $K=80$ and $-34\%$ at $K=120$, reflecting persistent misspecification bias that does not diminish with increasing $K$. In contrast, IJSE achieves relative errors of $-30\%$ at $K=80$ and $-19\%$ at $K=120$, and tracks the bootstrap closely (within 3 percentage points of relative error in both conditions). Coverage for IJSE and NP reaches 77--83\%, compared to 75--78\% for PostSD. IJSE runs in 0.08--0.15\,s per replication, while the cluster bootstrap requires 1.1--1.9\,s, roughly 12 to 14 times faster. The timing advantage over the bootstrap is more pronounced than in the earlier ICC-only simulation because the five-block Gibbs sampler (with the covariate update) imposes greater per-refit cost on the bootstrap, while the IJSE overhead remains dominated by the $O(KT)$ covariance computation.


\section{Simulation Study 4: $R^2$ in Multilevel Models}
\label{sec:sim_r2}

In addition to the ICC, applied researchers routinely report measures of explained variance in multilevel models. The marginal $R^2$ quantifies the proportion of total variance accounted for by fixed effects alone, while the conditional $R^2$ captures the proportion explained jointly by fixed effects and random effects \citep{NakagawaSchielzeth2013,JohnsonOmland2004}. Both measures are nonlinear ratios of variance components and are subject to the same misspecification concerns as the ICC. Because the data-generating process and working model are identical to those of Simulation Study~3, this section evaluates IJSE for the two $R^2$ functionals using the same MCMC output at no additional fitting cost.

\subsection{Target Functionals}
\label{subsec:r2_functionals}

Under the working model \eqref{eq:icc_working} with the fixed-effect covariate, let $\sigma_f^2 = \beta^2\,\mathrm{Var}(x)$ denote the variance of the fixed-effect predictions, where $\mathrm{Var}(x)$ is the empirical variance of the observed covariates. The marginal $R^2$ is defined as
\begin{equation}
g_5(\bm{\theta})
\equiv
R^2_m(\bm{\theta})
=
\frac{\sigma_f^2}{\sigma_f^2 + \sigma_U^2 + \sigma_\varepsilon^2},
\label{eq:r2m}
\end{equation}
and the conditional $R^2$ as
\begin{equation}
g_6(\bm{\theta})
\equiv
R^2_c(\bm{\theta})
=
\frac{\sigma_f^2 + \sigma_U^2}{\sigma_f^2 + \sigma_U^2 + \sigma_\varepsilon^2}.
\label{eq:r2c}
\end{equation}
At each posterior draw $t$, $\sigma_f^{2,(t)} = \beta^{(t),2}\,\mathrm{Var}(x)$, and both functionals are computed from $(\beta^{(t)}, \sigma_U^{2,(t)}, \sigma_\varepsilon^{2,(t)})$ using \eqref{eq:r2m}--\eqref{eq:r2c}. PostSD and IJSE are obtained directly from these draws and the cluster-level log-likelihood matrix \eqref{eq:icc_Lk} already computed for $g_4$. The bootstrap likewise reuses the same cluster resamples, computing the posterior means of $g_5$ and $g_6$ alongside $g_4$ from each bootstrap Gibbs run. All data-generating, fitting, and bootstrap details are identical to Section~\ref{subsec:icc_dgp}--\ref{subsec:icc_fitting}.

Structurally, $R^2_m$ places only the fixed-effect variance $\sigma_f^2 = \beta^2\,\mathrm{Var}(x)$ in the numerator, so it depends primarily on the regression coefficient $\beta$, whose posterior is relatively well identified even under misspecified error distributions. By contrast, $R^2_c$ includes the between-cluster variance $\sigma_U^2$ in the numerator, making it structurally analogous to the ICC $g_4 = \sigma_U^2/(\sigma_U^2 + \sigma_\varepsilon^2)$ and similarly vulnerable to the overly concentrated posteriors for variance parameters that arise under heavy-tailed data.

\subsection{Results}
\label{subsec:results_r2}

Table~\ref{tab:r2_results} presents the simulation results for the marginal and conditional $R^2$ under the same heavy-tailed heteroskedastic DGP as Simulation Study~3.

\begin{table}[!htb]
\centering
\caption{Simulation results for the marginal $R^2$ ($g_5$) and conditional $R^2$ ($g_6$) under heavy-tailed heteroskedastic errors ($R=300$ replications per $K$). Column definitions follow Table~\ref{tab:mediation_results}. Timings are shared with Table~\ref{tab:icc_results} because all three functionals ($g_4$, $g_5$, $g_6$) are computed from the same MCMC run.}
\label{tab:r2_results}
\small
\setlength{\tabcolsep}{3.5pt}
\begin{tabular}{cl cccrc cccrc}
\toprule
& & \multicolumn{5}{c}{Marginal $R^2$ ($g_5$)}
  & \multicolumn{5}{c}{Conditional $R^2$ ($g_6$)} \\
\cmidrule(lr){3-7} \cmidrule(lr){8-12}
$K$ & Method
  & $\mathrm{SE}_{\mathrm{MC}}$ & Mean SE & Bias & RelErr & EC
  & $\mathrm{SE}_{\mathrm{MC}}$ & Mean SE & Bias & RelErr & EC \\
\midrule
40 & PostSD & 0.042 & 0.037 & $-0.005$ & $-12\%$ & 0.90
            & 0.077 & 0.061 & $-0.017$ & $-22\%$ & 0.90 \\
   & IJSE   &       & 0.038 & $-0.005$ & $-11\%$ & 0.89
            &       & 0.062 & $-0.015$ & $-20\%$ & 0.89 \\
   & NP     &       & 0.038 & $-0.004$ & $-9\%$  & 0.90
            &       & 0.061 & $-0.017$ & $-21\%$ & 0.91 \\[2pt]
80 & PostSD & 0.030 & 0.026 & $-0.004$ & $-14\%$ & 0.90
            & 0.069 & 0.045 & $-0.023$ & $-34\%$ & 0.84 \\
   & IJSE   &       & 0.028 & $-0.003$ & $-9\%$  & 0.92
            &       & 0.053 & $-0.016$ & $-23\%$ & 0.87 \\
   & NP     &       & 0.028 & $-0.002$ & $-7\%$  & 0.92
            &       & 0.051 & $-0.017$ & $-25\%$ & 0.90 \\[2pt]
120 & PostSD & 0.025 & 0.022 & $-0.003$ & $-13\%$ & 0.93
             & 0.051 & 0.037 & $-0.014$ & $-27\%$ & 0.85 \\
    & IJSE   &       & 0.023 & $-0.002$ & $-8\%$  & 0.93
             &       & 0.044 & $-0.008$ & $-15\%$ & 0.88 \\
    & NP     &       & 0.023 & $-0.002$ & $-9\%$  & 0.92
             &       & 0.042 & $-0.009$ & $-18\%$ & 0.90 \\
\bottomrule
\end{tabular}
\end{table}

The two $R^2$ measures exhibit markedly different sensitivity to misspecification, consistent with their structural differences noted above. For the marginal $R^2$ ($g_5$), PostSD's relative error ranges from $-12\%$ to $-14\%$, considerably milder than the $-32\%$ to $-42\%$ observed for the ICC ($g_4$) in Table~\ref{tab:icc_results}. This attenuation reflects the fact that $R^2_m$ is dominated by $\beta^2\,\mathrm{Var}(x)$ in the numerator, and the posterior for the regression slope $\beta$ is relatively robust to misspecification of the error distribution. IJSE provides a modest further improvement, reducing relative error to $-8\%$ to $-11\%$, and tracks the bootstrap within 2 percentage points across all values of $K$. Coverage for all three methods ranges from 89\% to 93\%, substantially closer to the nominal 95\% than the coverage observed for the ICC.

For the conditional $R^2$ ($g_6$), the pattern more closely resembles the ICC. PostSD's relative error ranges from $-22\%$ at $K=40$ to $-34\%$ at $K=80$, reflecting the presence of $\sigma_U^2$ in the numerator and the same overly concentrated variance posteriors that drive ICC miscalibration. IJSE corrects a substantial portion of this bias, achieving relative errors of $-15\%$ to $-23\%$, and tracks the bootstrap within 3 percentage points. Coverage improves from 84--90\% under PostSD to 87--90\% under IJSE and NP.

The contrast between $g_5$ and $g_6$ illustrates a general principle: functionals whose numerators involve only fixed-effect parameters are less vulnerable to misspecification of the error distribution, while those that depend on random-effect variance components inherit the full severity of the Gaussian model's miscalibration. This distinction has practical implications for applied researchers reporting multilevel $R^2$ measures, because the conditional $R^2$, which is the more commonly reported of the two \citep{NakagawaSchielzeth2013}, is also the more susceptible to misspecification-driven underestimation of uncertainty.

\section{Discussion and Conclusion}
\label{sec:discussion}

This paper discussed and evaluated the infinitesimal jackknife standard error at both observation and cluster levels and evaluated it through four simulation studies covering six nonlinear functionals, which are often used in social and behavioral sciences: the unstandardized and standardized indirect effects ($g_1$, $g_2$), the ANOVA effect size $\eta^2$ ($g_3$), the intraclass correlation ($g_4$), and the marginal and conditional $R^2$ ($g_5$, $g_6$). Each study featured heavy-tailed, heteroskedastic data-generating processes that violate Gaussian working models, and compared IJSE against PostSD, which lacks robustness to misspecification, and the nonparametric bootstrap, which provides robustness but at substantially greater computational cost.

Across all four studies, PostSD consistently underestimated the frequentist standard error under misspecification, with relative errors ranging from $-12\%$ for the marginal $R^2$ to $-83\%$ for the mediation indirect effect at $N=1000$. Coverage fell as low as 57\%, far below the nominal 95\%. IJSE closely tracked the nonparametric bootstrap throughout, agreeing within 3 to 5 percentage points of relative error and producing comparable coverage, while running 3 to 28 times faster depending on the setting. Under correct specification (Simulation Study~1, Panel~A), all three methods agreed, confirming that IJSE introduces no distortion when the model is right.

The theoretical explanation for PostSD's failure is straightforward. Under misspecification, the posterior concentrates at a rate governed by the model-based Fisher information $H$, while the true sampling variability follows the sandwich form $H^{-1}JH^{-1}$, where $J$ reflects the actual score variance \citep{White1982,Muller2013}. Heavy tails and heteroskedasticity inflate $J$ relative to $H$, and this gap does not vanish with more data. Our simulations bear this out: in Simulation Study~2, PostSD's relative error held steady at $-33\%$ from $N=400$ to $N=600$, and in Simulation Study~3, it reached $-42\%$ at $K=80$. The severity of PostSD's miscalibration also depended on the functional. Functionals involving variance components in ratios or numerators, such as $\eta^2$, the ICC, the conditional $R^2$, and standardized coefficients, proved especially vulnerable because the Gaussian likelihood produces overly concentrated posteriors for variance parameters under heavy-tailed data \citep{GiordanoBroderick2023}. By contrast, the marginal $R^2$, whose numerator depends only on the fixed-effect coefficient $\beta$, exhibited the mildest misspecification impact (relative error $-12\%$ to $-14\%$), illustrating that functionals dominated by well-identified location parameters are more robust even when the error distribution is substantially misspecified.

The comparison between the ICC ($g_4$), marginal $R^2$ ($g_5$), and conditional $R^2$ ($g_6$) in Simulation Studies~3 and~4 is particularly informative, because all three functionals were computed from the same MCMC run under the same data-generating process. The progressive deterioration of PostSD from $g_5$ (mild) through $g_6$ (moderate) to $g_4$ (severe) provides a clean within-model demonstration that the degree of PostSD's miscalibration is governed by how much each functional depends on variance components rather than on location parameters.

These results support a clear practical recommendation. Because IJSE requires only the existing MCMC draws and per-observation (or per-cluster) log-likelihoods, it should be computed routinely alongside PostSD at negligible additional cost. When the two estimates agree, PostSD can be reported with confidence; when they diverge, the discrepancy serves as a diagnostic for misspecification, and IJSE should be preferred for constructing standard errors and confidence intervals. Researchers should, however, exercise caution when the number of independent units is small. In Simulation Study~3 with $K=40$ clusters, all three methods performed poorly, because the IJ approximation requires a sufficient number of independent contributions to stabilize the influence-function variance \citep{GiordanoBroderick2023}.

Several limitations should be noted. Our simulations used conjugate Bayesian models with Gibbs samplers; whether IJSE performs equally well with gradient-based samplers such as Hamiltonian Monte Carlo \citep{BrooksGelmanJonesMeng2011}, where MCMC autocorrelation may affect the covariance estimates in \eqref{eq:Ii_formula}, remains to be evaluated. The data-generating processes, though designed to reflect realistic departures from normality \citep{Micceri1989,Wilcox2012}, were stylized; other forms of misspecification including missing data, measurement error, and unobserved confounding were not considered. The coverage rates achieved by IJSE and the bootstrap, while consistently better than PostSD, remained below 95\% in most settings, because correct interval width does not guarantee correct coverage when the sampling distribution of $\bar{g}$ is itself non-Gaussian under severe misspecification \citep{DavisonHinkley1997}. We also did not include real-data applications, and practical value must ultimately be confirmed in empirical settings where the true data-generating process is unknown.

Looking ahead, evaluating IJSE for non-conjugate models and broader functional classes is a natural priority. \citet{JiLeeRabeHesketh2024} have already extended influence-function-based standard errors to Bayesian quantile regression with clustered data, and posterior quantiles, predictive probabilities, and other nonlinear summaries deserve similar investigation. The small-sample behavior of IJSE also warrants theoretical attention; higher-order corrections analogous to those for the bias-corrected bootstrap \citep{EfronTib1994} might improve performance when the number of independent units is limited. More broadly, understanding when and why the gap between model-based and sandwich variances is large for a given functional would help practitioners decide whether the sandwich correction is needed for their specific analysis \citep{KleijnVanDerVaart2012,Muller2013}.

In sum, IJSE provides a practical, computationally efficient, and theoretically grounded tool for uncertainty quantification in Bayesian workflows. It can be computed from a single MCMC run, closely approximates the nonparametric bootstrap, and exposes the miscalibration of PostSD under misspecification. We recommend that applied researchers adopt IJSE as a routine complement to PostSD, especially when the target estimand involves variance components or when the distributional assumptions of the working model may not hold.

\bibliography{cite}
\end{document}